\documentclass[prd,twocolumn,showpacs,superscriptaddress]{revtex4-1}

\usepackage{amsfonts}
\usepackage{amsmath}
\usepackage{amssymb}
\usepackage{bm}
\usepackage{dcolumn}
\usepackage{graphicx}
\usepackage{graphics}
\usepackage[latin1]{inputenc}
\usepackage{latexsym}
\usepackage{rotating}
\usepackage{xspace} 
\usepackage[usenames]{color}
\usepackage{mathrsfs}

\usepackage{ulem}
\definecolor {darkgreen}{rgb}{0.2,0.7,0.2}

\newcommand\be{\begin{equation}}
\newcommand\ba{\begin{eqnarray}}
\newcommand\ee{\end{equation}}
\newcommand\ea{\end{eqnarray}}

\newcommand\bw{\begin{widetext}}
\newcommand\ew{\end{widetext}}

\begin{document}
\title{I-Love-Q}

\author{Kent Yagi}
\affiliation{Department of Physics, Montana State University, Bozeman, MT 59717, USA.}

\author{Nicol\'as Yunes}
\affiliation{Department of Physics, Montana State University, Bozeman, MT 59717, USA.}

\date{\today}

\begin{abstract} 

Neutron stars and quark stars are not only characterized by their mass and radius, but also by how fast they spin, through their moment of inertia, and how much they can be deformed, through their Love number and quadrupole moment. 
These depend sensitively on the star's internal structure, and thus on unknown nuclear physics. 
We find universal relations between the moment of inertia, the Love number and the quadrupole moment that are independent of the neutron star's and quark star's internal structure. 
These can be used to learn about the deformability of these compact objects through observations of the moment of inertia, break degeneracies in gravitational wave detection to measure spin in binary inspirals and test General Relativity in a nuclear-structure independent fashion.

\end{abstract}



\maketitle

\emph{Introduction.}~One of largest uncertainties in nuclear physics is the relation between energy density and pressure at very high densities, the so-called equation of state (EoS). 
The interior structure of very compact stars, like neutron stars (NSs) and quark stars (QSs), depends sensitively on their EoS. This, in turn, also determines their exterior properties, such as their mass and radius, their rotation rate, characterized by their moment of inertia, and their deformability, characterized by their quadrupole moment and tidal Love number~\cite{Love,Postnikov}.
Surprisingly, however, we will show here that the moment of inertia, the quadrupole moment and the Love number satisfy a series of universal relations that are independent of the star's internal structure, provided one uses realistic EoSs.
We will also show how these relations have important implications to observational astrophysics, gravitational wave detections and experimental relativity. 

Some astrophysical observations allow us to infer properties of the EoS of compact stars~\cite{lattimer-prakash-review,ozel-review}. 
For example, the observation of X-ray bursters and low-mass X-ray binaries has allowed for the {\emph{simultaneous}} determination of the star's mass and radius to $\mathcal{O}(10)$\% accuracy~\cite{ozel-review}. 
Observations of double NS pulsars, such as J0737-3039~\cite{burgay}, may allow for the measurement of the moment-of-inertia to the same accuracy~\cite{lattimer-schutz,kramer-wex}. 
Gravitational wave (GW) observations from binary NS inspirals with second-generation ground-based detectors, such as Adv.~LIGO, Adv.~Virgo and KAGRA, may allow for the measurement of the tidal Love number~\cite{flanagan-hinderer-love,hinderer-lackey-lang-read,damour-nagar-villain}.    

None of these observations, however, are currently accurate enough to select between the many different EoSs that have been proposed, which then leads to degeneracies in the extraction of information from new observations. For example, GW observations of NS binary inspirals may have difficulty in extracting the individual spins, because these are degenerate with the quadrupole moment for non-precessing binaries\footnote{If the NS binary has misaligned spins, precession may break these degeneracies.}. Similarly, GWs from NS binary inspirals cannot be easily used to test General Relativity (GR), again due to degeneracies with the EoS~\cite{will-living,stairs}. 
We here find a way to uniquely break these degeneracies through universal \emph{I-Love-Q} relations, ie.~relations between the reduced moment-of-inertia $\bar{I}$, tidal Love number $\bar{\lambda}^\mathrm{(tid)}$ and quadrupole moment $\bar{Q}$, that are essentially insensitive to the star's EoS~\cite{Yagi:2013awa}. 

We find two possible reasons why these universal relations hold.
First, we find evidence that these quantities depend most sensitively on the star's interior structure close to its outer layer, precisely where our EoS ignorance is minimal and realistic EoSs agree. Second, we find evidence of a certain effacing of internal structure in the I-Love-Q relations; this is different from the GR effacement principle~\cite{damour-effacement}. As compactness is increased, we find that the I-Love-Q relations approach those of an isolated BH, as required by the no-hair theorems~\cite{hawking-uniqueness,carter-uniqueness}. 

The I-Love-Q relations have applications in several fields. 
On an observational astrophysics front, the measurement of a single member of the I-Love-Q trio would automatically provide information about the other two, even when the latter may not be accessible to observation. Such a measurement would provide all the information necessary to describe the exterior properties of tidally-deformed and slowly-rotating NSs and QSs at linear and quadratic order in spin. 
On a GW front, the I-Love-Q relations would break the degeneracy between the quadrupole moment and the spins in GWs emitted during binary NS inspirals. Given a GW detection from such a source with a second-generation ground based detector, the GW data analysis community may then be able to measure the averaged NS spin to about $0.01$ in dimensionless units.
On a fundamental physics front, the I-Love-Q relations will allow, for the first time, for tests of GR with NSs or QSs in the strong-field that are EoS independent.

\emph{Universal Relations.}~Consider an isolated, slowly-rotating NS or QS that is described by its mass $M_*$, the magnitude of its spin angular momentum $J$ and angular velocity $\Omega$, its (spin-induced) quadrupole moment $Q$ and its moment of inertia $I \equiv J/\Omega$.  Let us introduce dimensionless quantities $\bar{I} \equiv I/M_*^{3}$ and $\bar{Q} \equiv -Q/(M_*^{3} \chi^{2})$, where $\chi \equiv J/M_*^{2}$ is the dimensionless spin parameter\footnote{All throughout the paper we use geometric units with $G$ (Newton's gravitational constant) and $c$ (the speed of light) set to unity.}. The quantities introduced above have a clear physical meaning: $I$ determines how fast a body can spin given a fixed $J$; $Q$ encodes the amount of stellar quadrupolar deformation. These quantities are determined by solving the perturbed Einstein equations in a slow-rotation expansion ($\chi \ll 1$) to first and second order in spin, respectively~\cite{hartle1967,Yagi:2013awa}. Given a realistic EoS, such equations must be solved numerically. 

The slow-rotation approximation requires that $\chi$ be small enough such that all equations can be expanded in $\chi \ll 1$. In this approximation, the neglected corrections to the moment of inertia and quadrupole moment are of $\mathcal{O}(\chi^2)$ smaller than the leading-order contributions. Thus, demanding that any subleading terms be less than 10$\%$ of the leading-order ones forces the spin to satisfy $\chi \ll 0.3$, which corresponds to spin frequencies $\ll 600 \; \mathrm{Hz}$ or spin periods $\gg 1.7 \; \mathrm{ms}$. This implies that ``true'' millisecond pulsars, ie.~those with periods of $\sim 1 \; {\rm{ms}}$, cannot be modeled in a slow-rotation expansion. Double NS binary pulsars, however, are expected to be spinning much more slowly, and thus, the slow-rotation approximation would be adequate for them. 

In the presence of a companion, a NS or a QS will also be quadrupolarly deformed. The quadrupole moment tensor $Q_{ij}$ determines the magnitude of this deformation and it can be written as  $Q_{ij} = - \lambda^\mathrm{(tid)} \mathcal{E}_{ij}$, where $\lambda^{\mathrm{(tid)}}$ is the tidal Love number and $\mathcal{E}_{ij}$ is the quadrupole (gravitoelectric) tidal tensor that characterizes the source of the perturbation~\cite{flanagan-hinderer-love,hinderer-love}. Let us introduce the dimensionless tidal Love number $\bar{\lambda}^\mathrm{(tid)} = \lambda^\mathrm{(tid)}/M_*^5$, which physically characterizes the tidal deformability of a star in the presence of the companion's tidal field. $\bar{\lambda}^\mathrm{(tid)}$ can also be calculated by treating the tidal effect of the companion star as the perturbation to the isolated (non-rotating) NS or QS solution~\cite{hinderer-love,Yagi:2013awa}. 

We here present universal relations between $\bar{I}$, $\bar{Q}$ and $\bar{\lambda}^\mathrm{(tid)}$ for NSs and QSs that are essentially insensitive to their EoSs~\cite{Yagi:2013awa}. One might have expected these relations because $\bar{I} \propto C^{-2}$, $\bar{Q} \propto C^{-1}$ and $\bar{\lambda}^\mathrm{(tid)} \propto C^{-5}$ for polytropic EoSs in the Newtonian limit~\cite{Yagi:2013awa}, where $C = M_{*}/R_{*}$ is the compactness parameter, ie.~the ratio of the star's mass $M_*$ to its radius $R_*$. In the slow-rotation and small-deformation approximations, these barred quantities depend on spin only quadratically, and thus, for slowly-rotating stars, the relations are essentially spin-independent.

We consider 6 different realistic EoSs for NSs: APR~\cite{APR}, SLy~\cite{SLy}, Lattimer-Swesty with nuclear incompressibility of 220MeV (LS220)~\cite{LS}, Shen~\cite{Shen1}, PS~\cite{PS} and PCL2~\cite{SQM}, and a simple $n=1$ polytropic EoS, with $p=K \rho^{1+1/n}$. For the LS220 and Shen EoSs, we adopt a temperature of $0.1 \; {\rm{MeV}}$ and assume they are neutrino-less and in $\beta$-equilibrium. For QSs, we consider 3 EoSs: SQM1, SQM2 and SQM3~\cite{SQM}. We assume the stars are uniformly rotating, with isotropic pressure.

%
\begin{figure*}[htb]
\begin{center}
\begin{tabular}{l}
\includegraphics[width=7.8cm,clip=true]{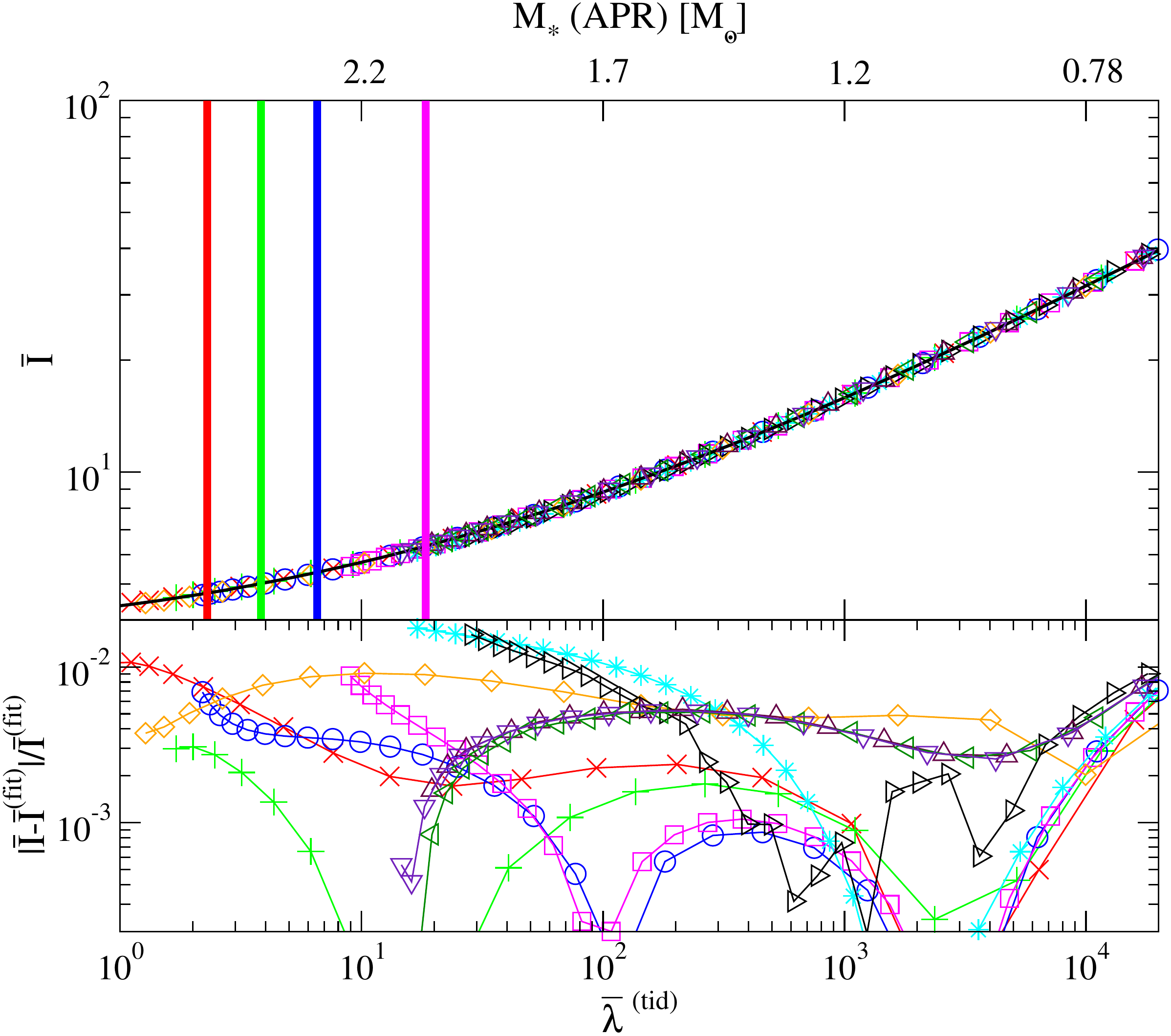} 
\includegraphics[width=7.8cm,clip=true]{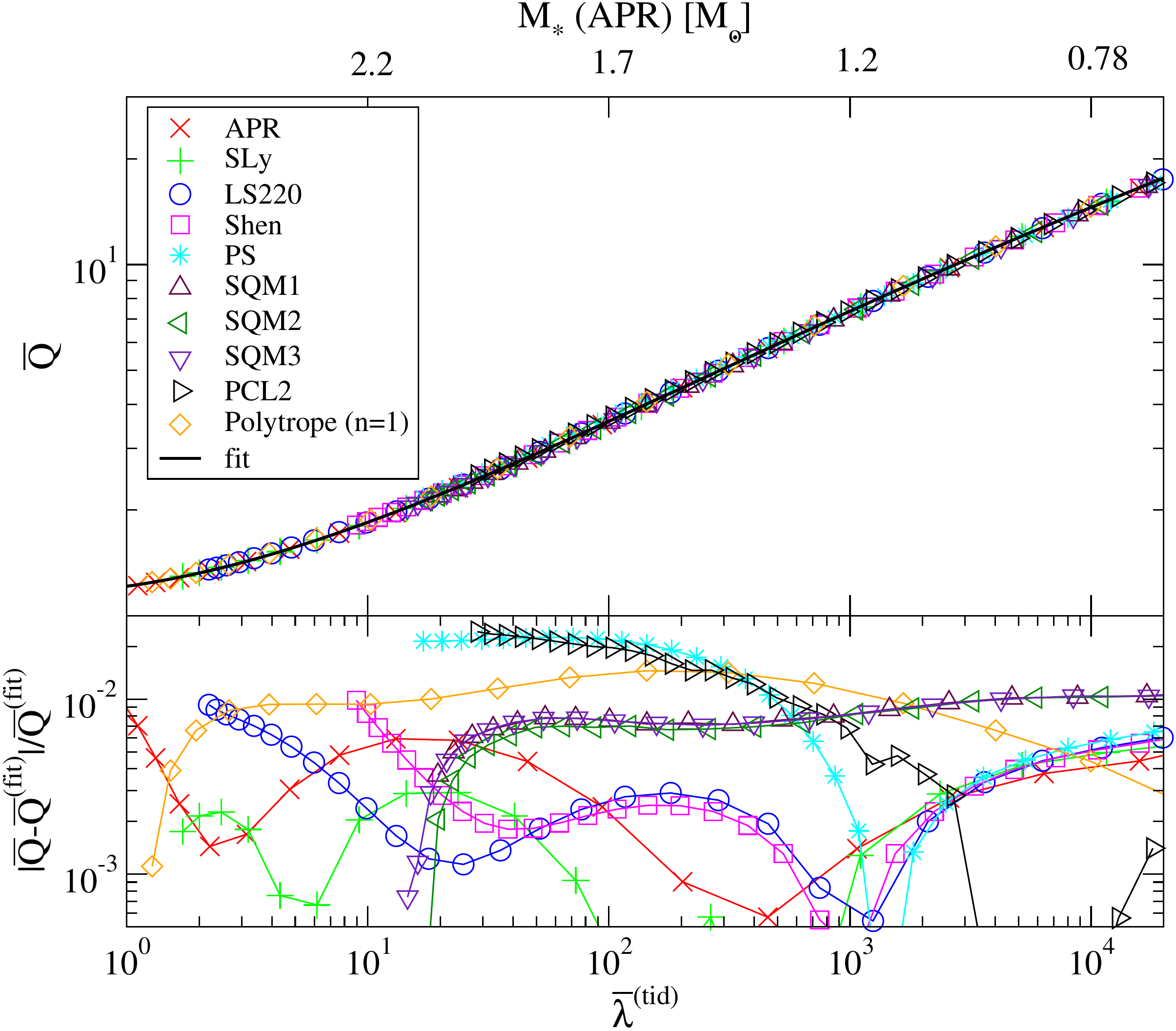} 
\end{tabular}
\caption{
(Top Left and Right) The neutron star (NS) and quark star (QS) universal I-Love and Love-Q relations for various EoSs, together with fitting curves (solid). On the top axis, we show the corresponding NS mass with an APR EoS. The thick vertical lines show the stability boundary for the APS, SLy, LS220 and Shen EoSs from left to right. The parameter varied along each curve is the NS or QS central density, or equivalently the star's compactness, with the latter increasing to the left of the plots. (Bottom Left and Right) Fractional errors between the fitting curve and numerical results.
}
\end{center}
\end{figure*}
{\renewcommand{\arraystretch}{1.2}
\begin{table}
\begin{centering}
\begin{tabular}{cccccccc}
\hline
\hline
\noalign{\smallskip}
 $y_i$ & $x_i$ &&  \multicolumn{1}{c}{$a_i$} &  \multicolumn{1}{c}{$b_i$}
&  \multicolumn{1}{c}{$c_i$} &  \multicolumn{1}{c}{$d_i$} &  \multicolumn{1}{c}{$e_i$}  \\
\hline
\noalign{\smallskip}
 $\bar{I}$ & $\bar{\lambda}^{\rm{(tid)}}$ && 1.47 & 0.0817  & 0.0149 & $2.87\times 10^{-4}$ & $-3.64\times 10^{-5}$\\
 $\bar{I}$ & $\bar{Q}$ && 1.35  & 0.697 & -0.143  & $9.94\times 10^{-2}$ & $-1.24\times 10^{-2}$\\
 ${\bar{Q}}$ & $\bar{\lambda}^{\rm{(tid)}}$ && 0.194  & 0.0936 & 0.0474  & $-4.21\times 10^{-3}$ & $1.23\times 10^{-4}$\\
\noalign{\smallskip}
\hline
\hline
\end{tabular}
\end{centering}
\caption{Estimated numerical coefficients for the fitting formulas of the NS and QS I-Love, I-Q and Love-Q relations.}
\end{table}
}
The top panels of Fig.~1 present the NS and QS I-Love and Love-Q relations for various EoSs. Observe that these relations hold universally for both NS and QS sequences, essentially independently of their EoSs~\footnote{In a previous version of this paper~\cite{Yagi:2013bca}, we had found that the universal I-Love-Q relations for NSs and QSs separated into two distinct branches. This was an error associated with sensitive boundary conditions for QSs. Upon correcting this, the QS branch joins the NS one and both stars share the same universality.}. A similar universal relation holds between $\bar{I}$ and $\bar{Q}$~\cite{Yagi:2013awa}. Such relations can be numerically fitted with a polynomial on a log-log scale~\cite{Yagi:2013awa}, shown in Fig.~1 with solid and dashed black curves, namely:  
\begin{equation}
\ln y_i = a_i + b_i \ln x_i + c_i (\ln x_i)^2 + d_i (\ln x_i)^3+ e_i (\ln x_i)^4\,,
\label{fit}
\end{equation}
where the coefficients are summarized in Table~1. The bottom panels of this figure show the fractional errors between the fitted curves and the numerical results. Equation~(\ref{fit}) is a numerical fit, because the data in Fig.~1 is itself obtained by numerically solving the Einstein structure equations, which in turn is unavoidable for realistic EoSs. For very simple polytropic EoSs, where the equations of structure can be solved analytically in the Newtonian limit, one can obtain similar universal relations that are purely analytic~\cite{Yagi:2013awa}. 

We have found two possible reasons that may explain the I-Love-Q relations. First, we find some evidence that the mathematical relations that define $I$, $\lambda^{(\rm{tid})}$ and $Q$ depend mostly on the star's internal structure near its outer layer, where our ignorance of nuclear physics is minimal and realistic EoSs agree. For example, the integral that defines $I$ in the Newtonian limit accumulates the most near the surface~\cite{Yagi:2013awa}. This evidence then suggests that the I-Love-Q scaling relations should lose their universality for unrealistic EoSs that modify the star's internal structure near its surface. We have verified this explicitly by computing these relations for NSs with $n=2$, $2.5$ and $3$ polytropic EoSs: the I-Love-Q curves deviate away from those in Fig.~1 as $n$ increases. 

The second reason is related to the no-hair theorems of GR. Figure~1 shows that the I-Love-Q relations approach the expected I-Love-Q relations for BHs, ie.~$\bar{I} \to 4$, $\bar{\lambda}^\mathrm{(tid)} \to 0$ and $\bar{Q} \to 1$~\cite{Yagi:2013awa}. For BHs, all multipole moments of the exterior spacetime are related to the BH mass and spin~\cite{geroch,hansen} (e.g.~there is a well-known I-Q relation) because of the no-hair theorems~\cite{hawking-uniqueness,carter-uniqueness}. But for NSs and QSs, such relations were thought to not hold due to the lack of no-hair theorems for non-vacuum spacetimes. In spite of this, our results suggest the existence of NS and QS universal relations between $I$ and $Q$ that are similar to those that arise for BHs, and perhaps, hint at the existence of something similar to a no-hair theorem for non-vacuum spacetimes. 

The I-Love-Q scaling found here suggests an {\emph{effacing of internal structure}}, i.e. the expected internal-structure dependence of the I-Love-Q trio is effectively not there. This is not a consequence of the well-known {\emph{effacement principle}}~\cite{damour-effacement} in GR, as the latter applies only to the motion of BHs. The I-Love-Q relations found here relate different multipole components of the exterior gravitational field of isolated bodies and says nothing about their relative motion.

\emph{Application to Observational Astrophysics.}~Double NS binary pulsars have the potential to measure $I$ with $10\%$ accuracy in the near future~\cite{lattimer-schutz,kramer-wex}. The moment of inertia may be measurable because it induces additional periastron precession, as well as precession of the angular momentum vector and the NS spin vectors. The precession of the former translates into a time-dependent inclination angle, while the precession of the latter may force the pulsar beams to sweep in and out of Earth's line of sight. Alternatively, this precession may only cause a change in the observed average pulse shape, as is the case for the Hulse-Taylor binary pulsar, in which case direct measurement may be more difficult. 

Given an observed $I_{\rm obs}$, $M_{\rm obs}$ and $\Omega_{\rm obs}$, the I-Love-Q relations automatically provide the value of $\lambda^{(\rm{tid})}$ and $Q$. These two quantities would not be easily observable with binary pulsars directly; although $Q$ and $\lambda^{(\rm{tid})}$ do induce additional precession, their effect is suppressed relative to that of $I$, by various powers of the ratio between the binary's orbital velocity and the speed of light. Of course, the I-Love-Q relations refer to reduced (barred) quantities, which must be appropriately normalized by the mass and spin period. The former differs from the observed mass by quantities of ${\cal{O}}(\chi^{2})$, and a small error in the observed mass could induce a large error in derived quantities. Such an error is smaller than the non-universality of the I-Love-Q relations if the NS spin period is much greater than 8.5ms, which is the case for the double pulsar binary and NS binaries in the LIGO band.

\emph{Application to GW Astrophysics.}~Interferometric GW detectors are most sensitive to the GW phase of the signal. For waves emitted during NS binary inspirals, the GW phase contains a term proportional to the NSs' spin-induced quadrupole moments, $Q_{1}$ and $Q_{2}$, and another term proportional to their tidally-induced quadrupolar deformations $\lambda_{1}^{(\mathrm{tid})}$ and $\lambda_{2}^{(\mathrm{tid})}$. The former enters with a factor proportional to $v^{4}/c^{4}$~\cite{poisson-quadrupole}, while the latter is proportional to $v^{10}/c^{10}$~\cite{flanagan-hinderer-love}, relative to the leading-order term. Since by Kepler's third law, the orbital velocity is related to the GW frequency via $v \propto f^{1/3}$, each term has a distinct frequency dependence that in principle makes them non-degenerate. 

The NS quadrupole moment is degenerate with the NSs' individual spins, because there is a spin-spin interaction term in the GW phase that enters at the same order in $v/c$ as the quadrupole one~\cite{poisson-quadrupole}. Such a degeneracy may prevent us from simultaneously extracting the quadrupole moment and the individual spins from a GW detection. The Love-Q relation, however, can be used to break this degeneracy, by rewriting $\bar{Q}$ as a function of $\bar{\lambda}^\mathrm{(tid)}$. If the Love number can be measured with a GW detection, then one can also separately measure the spins.

Let us investigate the impact of the Love-Q relation on NS spin determination, by comparing parameter estimation with and without this relation. When using the Love-Q relation, we parameterize the spin-dependent part of the waveform phase with the dimensionless averaged spin $\chi_s$ and the dimensionless spin difference $\chi_a$. When not using the Love-Q relation, we parameterize the phase with the effective spin-parameter $\beta$, constructed from a certain combination of the individual spins, which is the leading spin-contribution to the phase. We perform a Fisher analysis to calculate the measurement accuracy of such spin parameters, taking correlations with other parameters (such as NS masses) into account. 

\begin{figure}[htb]
\begin{center}
\includegraphics[width=7.8cm,clip=true]{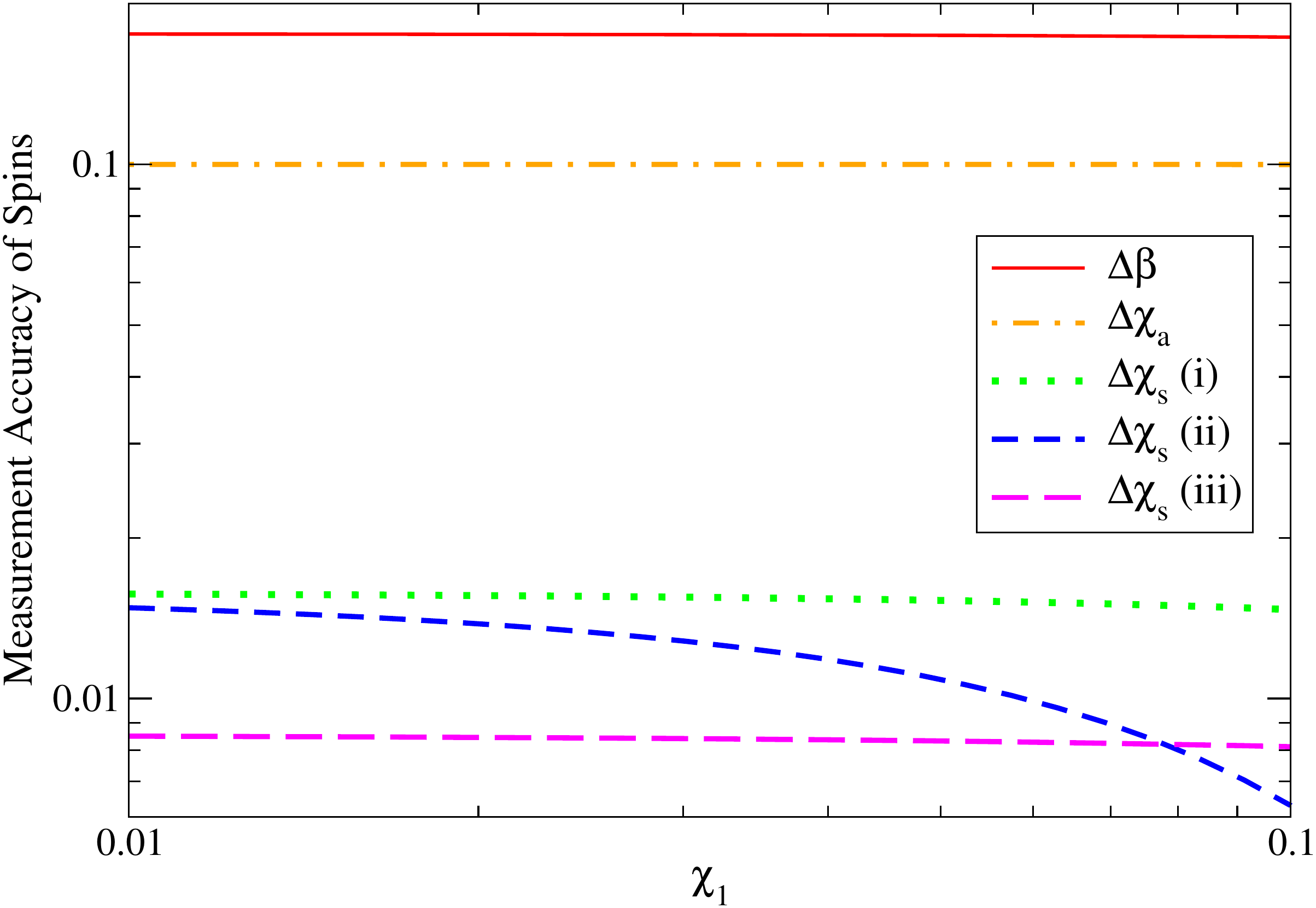}  
\caption{
Measurement accuracy of spin parameters $\beta$, $\chi_s$ and $\chi_a$ with Adv.~LIGO given a detection at $D_{L} =100$ Mpc with $\mathrm{SNR} \approx 30$. $\beta$ is an effective spin parameter, constructed from the individual spins, while $\chi_{s}$ and $\chi_{a}$ are the dimensionless spin average and difference respectively. We consider three different NS binaries [(i), (ii) and (iii)] as described in the text. $\Delta \beta$ is computed without using the NS Love-Q relation, while $\Delta \chi_{s,a}$ are computed using this relation. $\Delta \beta$ and $\Delta \chi_a$ are almost identical for all systems, since they are dominated by their priors ($|\beta|<0.2$ and $|\chi_a|<0.1$). Thanks to the Love-Q relation, $\chi_{s}$ can be measured to $\sim 0.01$.}
\end{center}
\end{figure}
Figure~2 shows the measurement accuracy of $\chi_{s}$, $\chi_{a}$ and $\beta$, given a binary NS observation at $D_{L}=100$ Mpc with Adv.~LIGO and a signal-to-noise ratio (SNR) of $\approx 30$ for three different systems: (i) $(m_1,m_2)=(1.45,1.35)M_\odot$, $\chi_1=\chi_2$, (ii) $(m_1,m_2)=(1.45,1.35)M_\odot$, $\chi_1=2 \chi_2$ and (iii) $(m_1,m_2)=(1.4,1.35)M_\odot$, $\chi_1=\chi_2$. The solid red line corresponds to the measurement accuracy of $\beta$ without the NS Love-Q relation, while the other lines correspond to the measurement accuracy of $\chi_{s}$ and $\chi_{a}$ with the NS Love-Q relation. Observe that $\Delta \beta$ and $\Delta \chi_a$ are almost identical for all systems, since they are dominated by their priors. On the other hand, thanks to the Love-Q relation, one can determine the averaged spin $\chi_s$ to $\sim0.01$.
s
\begin{figure*}[thb]
\begin{center}
\begin{tabular}{l}
\includegraphics[width=8.75cm,clip=true]{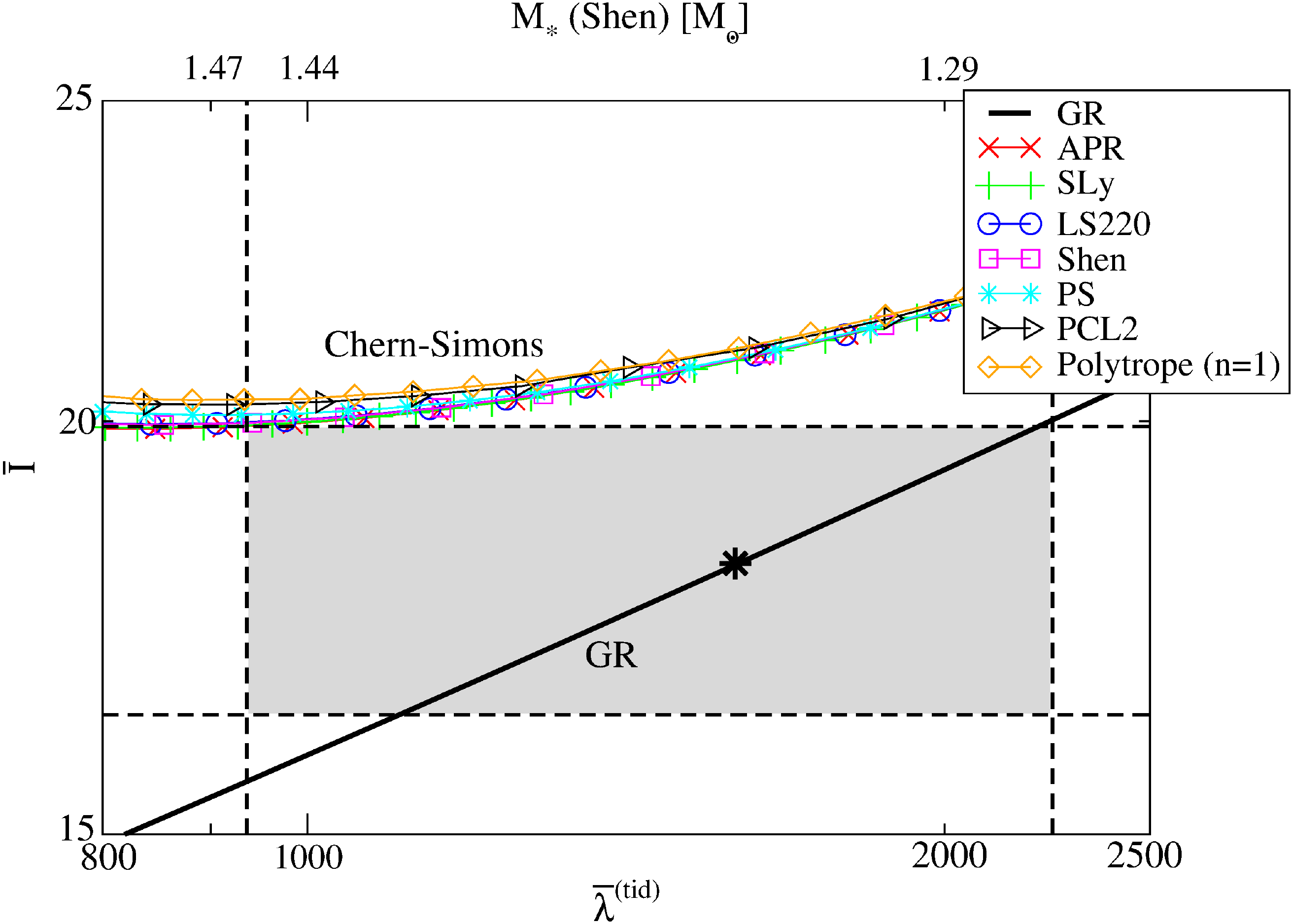}  
\includegraphics[width=8.75cm,clip=true]{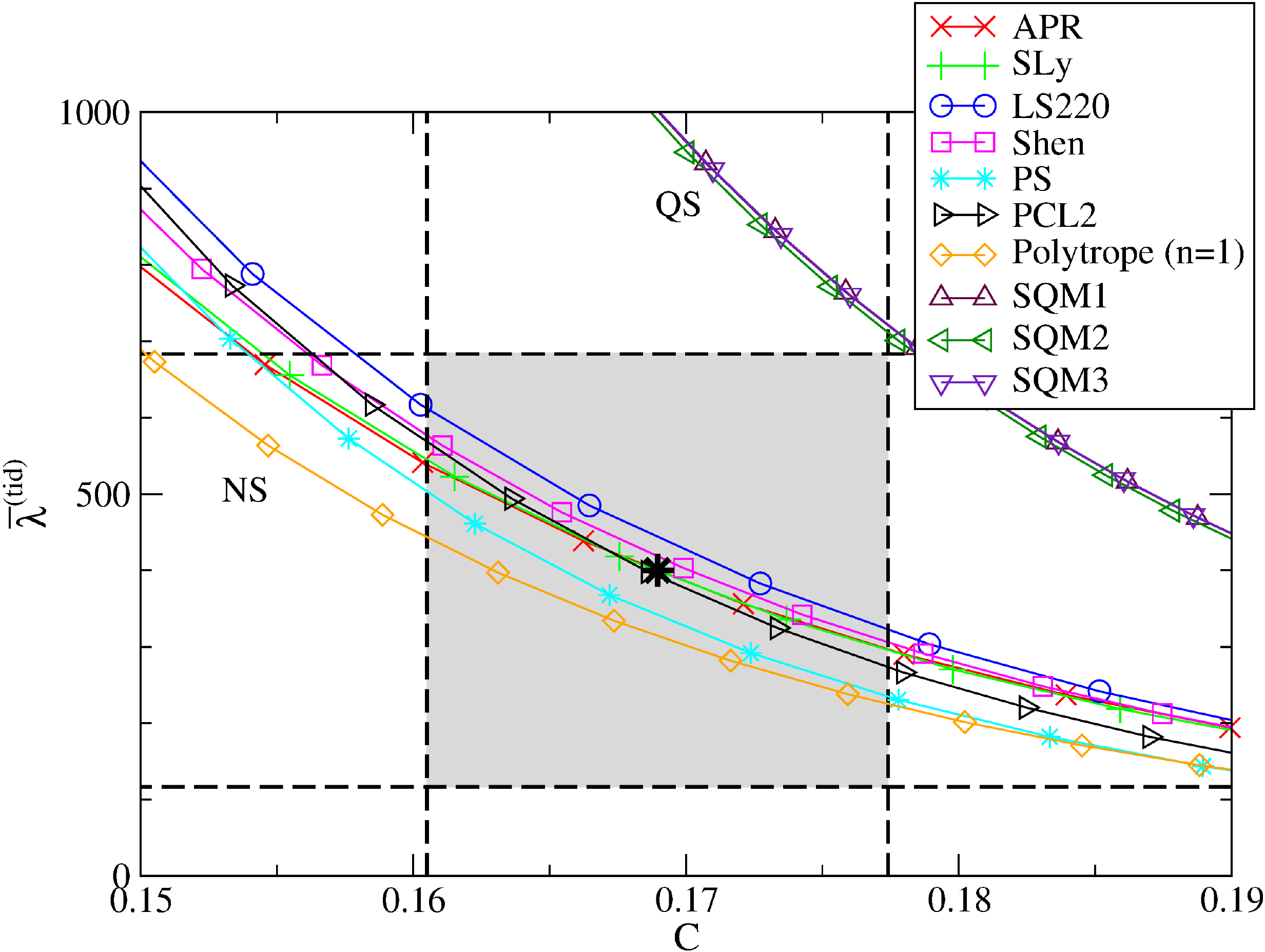}  
\end{tabular}
\caption{
(Left) Possible error box (shaded region) in the I-Love plane, given two independent observations of the moment of inertia and the tidal Love number, shown with a black star. The black solid line shows the NS and QS I-Love relation in GR, while all the other lines show the same relation in dynamical Chern-Simons (CS) gravity. These measurements would force the CS coupling parameter to be small enough that its I-Love relation falls inside of the error box, leading to a constraint that is 6 orders of magnitude stronger than current Solar System ones. The top axis show the NS mass for the Shen EoS.
(Right) Possible error box (shaded region) in the Love-C plane, given two independent observations of the Love number and the NS compactness, shown with a black star. The different curves show the Love-C relation in GR for several EoSs. Observe that the difference between these curves for NSs is smaller than the error box, thus allowing for a generic test of GR. Such a test, however, requires the assumption that the observed object is a NS and not a QS, since the latter has a different Love-C relation.}
\end{center}
\end{figure*}

\emph{Application to Fundamental Physics.}~Pulsar observations allow for tests of GR~\cite{stairs} in regions of the Universe where the gravitational field is much stronger than in the Solar System~\cite{will-living}. Unfortunately, however, these tests are not effective most of the time because of degeneracies between modified gravity effects and the EoS. The I-Love-Q relations can be used to break this degeneracy and thus allow for EoS-independent tests of GR.

A robust GR test would require at least two independent measurements of any two quantities in the I-Love-Q trio. Given a single measurement of any one of them, the I-Love-Q relations give us what the other two values must be for a NS or a QS in GR. A second independent measurement can then be used as a redundancy test, irrespective of the EoS or whether the star is a NS or a QS. If the two observables do not lie on the I-Love-Q line as predicted in GR, then the observations would indicate a GR deviation. Of course, such tests are effective, provided these measurements have a sufficiently small uncertainty.    

As an example, let us assume that a binary pulsar observation has led to a measurement of $\bar{I}$ to $10\%$ accuracy~\cite{lattimer-schutz,kramer-wex}, while a GW observation has led to a measurement of $\bar{\lambda}^\mathrm{(tid)}$ to roughly $60\%$ for a system with similar masses~\cite{flanagan-hinderer-love,hinderer-lackey-lang-read,damour-nagar-villain,Yagi:2013awa}. Such a measurement would lead to an error box in the I-Love plane, centered about the measured value, as shown by the shaded region in the left panel of Fig.~3. In this example, such an observation would be consistent with the GR I-Love relation for NSs and QSs. Any modified gravity theory is then constrained to predict an I-Love relation that runs through this error box. If the modified gravity theory predicts a modified I-Love-Q relation, as non-trivial theories do, this I-Love test would constrain the coupling constants of the theory at scales comparable to the NS's or QS's radius.  

Let us consider a specific modified gravity theory: dynamical Chern-Simons (CS) gravity~\cite{CSreview}. This theory modifies Einstein's by introducing a gravitational parity-violating interaction through a dynamical scalar field and it is currently constrained only very weakly by known tests and experiments~\cite{CSreview}. Figure~3 shows the CS I-Love curves for NSs with a variety of EoSs and a specific value of the CS coupling constant. Observe that these curves are essentially insensitive to the NS EoS. Observe also that the CS I-Love relation for NSs lie above that in GR; this is true regardless of the value of the CS coupling constant. Therefore, in this example, the hypothetical measurement and error box of Fig.~3 would force the CS coupling parameter to be small enough for the CS NS I-Love curve to go through the error box. This would lead to a stringent constraint on the theory, one that is $\sim 6$ orders of magnitude stronger~\cite{Yagi:2013awa} than current Solar System tests~\cite{alihaimoud-chen}. 

Even if one were not able to measure the moment of inertia in the near future, one can still test GR through the relation between the tidal Love number and the compactness $C$, as originally found by~\cite{Postnikov,hinderer-lackey-lang-read}. Let us then assume again that a GW observation has measured $\bar{\lambda}^\mathrm{(tid)}$ to roughly $60\%$~\cite{Yagi:2013awa} and that future low-mass X-ray binary observations have measured $C$ to $5\%$ for a system with similar masses~\cite{ozel-review}. Such observations would lead to an error box in the Love-C plane, centered about the observed value, as shown by the shaded region in the right panel of Fig.~3. The NS Love-C relation, however, is dependent on the EoS, as shown by the spread in curves. For NSs, the vertical distance between all curves is much smaller than the error in the measurement of the tidal Love number, making the NS Love-C relation {\emph{effectively}} EoS-independent. The requirement that any Love-C relation goes through this error box could constitute an effectively EoS independent GR test, although not as EoS independent as an I-Love-Q test. Such a test, however, requires the assumption that the object observed is a NS and not a QS, since the Love-C curves are quite different for these two objects, as shown on the right panel of Fig.~3. 

\emph{Discussions.}~The I-Love-Q relations open the door to exciting applications in astrophysics, GW theory and fundamental physics. We have here performed a cursory study of possible applications, but these could be followed up by much more detailed analysis. For example, the measurement accuracy of GW phase parameters was here estimated via a Fisher analysis, but this could be improved through Bayesian methods~\cite{cornish-PPE}. One could also extend these tests to GW and binary pulsar systems that do not have exactly the same masses. We have indeed verified that all the applications discussed above are robust, even when all the NS masses (those measured with binary pulsars and those measured with GWs) differ by about $10\%$~\cite{Yagi:2013awa}.

The analysis of the I-Love-Q relations presented here opens up the road for multiple follow-up studies. For example, one could determine whether these relations hold for NSs and QSs with anisotropic pressure~\cite{doneva}, large internal magnetic fields and rapid rotation~\cite{Berti:2004ny}. The latter may be particularly important, as differential rotation can produce stars with $J/M_*^2>1$~\cite{Giacomazzo:2011cv}, while the fastest known millisecond pulsars can have spin parameters as large as $0.5$, for which a quadratic slow-rotation expansion might not suffice. In fact, relativistic calculations of stellar structure for uniform, very fast rotation show that the quadrupole moment can differ by $\sim 20\%$ from the predictions of the slow-rotation approximation~\cite{Berti:2004ny}. Fortunately, however, those binary pulsars where the moment of inertia may be detected first will probably consist of slowly-rotating double NSs. These NSs rotate much more slowly than ``true'' millisecond pulsars, and thus, the slow-rotation approximation would be appropriate. Nevertheless, one could study the rapidly-rotating case by carrying out a fourth-order expansion in slow-rotation (or a numerical study~\cite{Berti:2004ny}) and investigating whether higher-order spin terms spoil the universal relations. One could also investigate whether there are universal relations between other quantities, such as the f- and w-modes of NS oscillations~\cite{andersson-kokkotas,tsui-leung}; even if the latter are difficult to observe, such universality may be interesting on theoretical grounds.

\emph{Acknowledgments}:
The authors thank Evan O'Connor and Benjamin Lackey for providing tabulated EoSs, as well as Emanuele Berti, 
Luc Blanchet, Vitor Cardoso, Tanja Hinderer, Kenta Hotokezaka, Michael Kramer, Lee Lindblom, Hiroyuki Nakano, Feryal $\ddot{\mathrm{O}}$zel, Paolo Pani, Eric Poisson, Scott Ransom, Luciano Rezzolla, Masaru Shibata, Takahiro Tanaka and the 4 anonymous reviewers for their comments. The authors also thank the Yukawa Institute for Theoretical Physics at Kyoto University, where this work was initiated during the Long-term Workshop YITP-T-12-03 on ``Gravity and Cosmology 2012''. NY acknowledges support from NSF grant PHY-1114374, as well as support provided by the National Aeronautics and Space Administration from grant NNX11AI49G, under sub-award 00001944. 
\bibliography{master}
\end{document}